\documentclass{article}

\newtheorem{theorem}{Theorem}

\input{tcilatex}

\begin{document}

\title{Critical Behaviour in a Planar Dynamical Triangulation Model with a Boundary}
\author{V. Malyshev \\
INRIA (France)}
\maketitle

\begin{abstract}
We consider a canonical ensemble of dynamical triangulations of a
2-dimensional sphere with a hole where the number $N$ of triangles is fixed.
The Gibbs factor is $\exp (-\mu \sum \deg v)$ where $\deg v$ is the degree
of the vertex $v$ in the triangulation $T$. Rigorous proof is presented that
the free energy has one singularity, and the behaviour of the length $m$ of
the boundary undergoes 3 phases: subcritical $m=O(1)$, supercritical
(elongated) with $m$ of order $N$ and critical with $m=O(\sqrt{N})$. In the
critical point the distribution of $m$ strongly depends on whether the
boundary is provided with the coordinate system or not. In the first case $m$
is of order $\sqrt{N}$, in the second case $m$ can have order $N^{\alpha }$
for any $0<\alpha <\frac{1}{2}$.
\end{abstract}

Dynamical triangulations is a popular approach to quantum gravity, see for
example \cite{amdujo, bmdgpd}. We consider here two-dimensional planar model
with the action, used earlier in \cite{kastwy}. The main peculiarity of our
paper is that the sphere contains a hole, the boundary length of which is a
dynamical variable. Our results show the existence of a critical point in
the canonical ensemble, three different phases and instability of
fluctuations of the boundary length at the critical point.

Many of our results do not depend on the class of triangulations considered
but for definiteness we consider triangulations $T$ of a two-dimensional
sphere with a hole, called quasitriangulations in \cite{goujac}. They
consist of vertices, edges and triangles, but are not assumed to be an
abstract simplicial complex. For example, several edges can connect two
vertices. Let $V(T)$ ($L(T)$) be the set of vertices (edges) in $T$, $L(T)$
includes the boundary edges.

Let $T_{0}(N,m)$ be the set of all such triangulations with $N$ triangles
and $m\geq 2$ edges on the boundary. We assume also that on the boundary one
vertex and an edge incident to it are specified, thus fixing the origin of
the coordinate system and the orientation. In $T_{0}(N,m)$ only equivalence
classes of such triangulations are counted, under homeomorphisms that
respect the origin and orientation. Let $C_{0}(T,m)$ be the number of such
equivalence classes. Canonical distribution is defined by the probability $%
P_{0,N}(T)$ of triangulation $T$ 
\begin{equation}
P_{0,N}(T)=Z_{0,N}^{-1}\exp (-\frac{\mu _{1}}{2}\sum_{v\in V(T)}\deg v
)
\end{equation}
This is a particular case (with parameters $t_{q}=t)$ of the Gibbs factor 
\begin{equation}
\prod_{q>2}t_{q}^{n(q,T)}
\end{equation}
used in \cite{kastwy}, where $n(q,T)$ is the number of vertices of $T$
having degree $q$. The partition function can be written as 
\begin{equation}
Z_{0,N}=\sum_{m}\sum_{T\in T_{0}(N,m)}\exp (-\frac{\mu _{1}}{2}\sum_{v}\deg
v)=\sum_{m}\sum_{T\in T_{0}(N,m)}\exp (-\mu _{1}\left| L(T)\right| )
\end{equation}
We will observe phase transitions with respect to parameter $\mu _{1}$. It
has the critical point $\mu _{1,cr}=\log 12$. Let $\beta _{0}=\beta _{0}(\mu
_{1})$ be such that 
\begin{equation}
\frac{(1+\frac{4\beta _{0}}{3(1-\beta _{0})})}{(1+\frac{2\beta _{0}}{1-\beta
_{0}})^{2}}\exp (-\mu _{1}+\log 12)=1
\end{equation}

\begin{theorem}
The free energy $\lim_{N}\frac{1}{N}\log Z_{0,N}=F$ is equal to $-\frac{3}{2}%
\mu _{1}+c,c=3\sqrt{\frac{3}{2}}$, if $\mu _{1}>\mu _{1,cr}$ and to 
\begin{equation}
-\frac{3}{2}\mu _{1}+c+\beta _{0}(-\mu _{1}+\log 12)+\int_{0}^{\beta
_{0}}\log (\frac{(1+\frac{4\beta }{3(1-\beta )})}{(1+\frac{2\beta }{1-\beta }%
)^{2}}d\beta 
\end{equation}
\ if $\mu _{1}<\mu _{1,cr}$.
\end{theorem}

Note that if $\mu _{1}\rightarrow \mu _{1,cr}$ then $\beta _{0}\rightarrow 0$%
.

Let $m(N)$ be the random length of the boundary when $N$ is fixed. Its
probability can be written as, using $\left| L(T)\right| =\frac{3N}{2}+\frac{%
m}{2}$,

\begin{equation}
P_{0,N}(m(N)=m)=\Theta _{0,N}^{-1}\exp (-\mu _{1}\frac{m}{2}%
)C_{0}(N,m),\Theta _{0,N}=\sum_{m}\exp (-\mu _{1}\frac{m}{2})C_{0}(N,m)
\end{equation}

\begin{theorem}
There are 3 phases, where the distribution of $m(N)$ has quite different
asymptotical behaviour:

\begin{itemize}
\item  Subcritical region, that is $12\exp (-\mu _{1})<1$: $m(N)=O(1)$, more
exactly the distribution of $m(N)$ has a limit $\lim_{N}P_{N}(m(N)=m)=p_{m}$
for fixed $m$ as $N\rightarrow \infty $. Thus the hole becomes neglectable
with respect to $N$.

\item  Supercritical region (elongated phase), that is $12\exp (-\mu _{1})>1$%
. Here the boundary length is of order $O(N)$. More exactly there exists $%
\varepsilon >0$ such that $\lim P_{0,N}(\frac{m_{N}}{N}>\varepsilon )=1$.

\item  In the critical point, that is when $12\exp (-\mu _{1})=1$, the
boundary length is of order $\sqrt{N}$. The exact statement is that the
distribution of $\frac{m_{N}}{\sqrt{N}}$ converges in probability.
\end{itemize}
\end{theorem}

Proof. We use the combinatorial method of Tutte \cite{tutte1} instead of
random matrix representation of two-dimensional gravity \cite{britpazu}. We
use the following formula \cite{goujac}, where $N=m+2j$, 
\begin{equation}
C_{0}(N,m)=\frac{2^{j+2}(2m+3j-1)!(2m-3)!}{(j+1)!(2m+2j)!((m-2)!)^{2}}
\end{equation}
By direct calculation we get, taking into account that $N\rightarrow
N,m\rightarrow m+2$ corresponds to $j\rightarrow j-1,m\rightarrow m+2$ 
\begin{equation}
\frac{P_{0,N}(m+2)}{P_{0,N}(m)}=f(N,m)=\frac{C_{0}(N,m+2)\exp (-\frac{\mu
_{1}}{2}(m+2))}{C_{0}(N,m)\exp (-\frac{\mu _{1}}{2}m)}=
\end{equation}
\begin{equation}
=\exp (-\mu _{1}+\log 12)\frac{(1+\frac{2}{N-m})(1+\frac{4m}{3(N-m)})}{(1+%
\frac{2m}{N-m}+\frac{2}{N-m})(1+\frac{2m}{N-m}+\frac{1}{N-m})}\frac{(1-\frac{%
1}{4m^{2}})}{(1-\frac{1}{m})}  \label{main}
\end{equation}

In the subcritical case for fixed $m$ and $N\rightarrow \infty $ 
\begin{equation}
\frac{P_{0,N}(m+2)}{P_{0,N}(m)}\sim \exp (-\mu _{1}+\log 12)(1+\frac{1}{m}+O(%
\frac{1}{m^{2}}))
\end{equation}
and thus as $m\rightarrow \infty $, for example for even length, $%
\lim_{N}P_{0,N}(2m)\sim _{m\rightarrow \in }Cm\exp (m(-\mu _{1}+\log 12))$.

At the same time the second factor in (\ref{main})\ is less than $1$. Thus
from 
\begin{equation}
Z_{0,N}=\exp (-\mu _{1}\frac{3N}{2})\Theta _{0,N}=\exp (-\mu _{1}\frac{3N}{2}%
)\sum_{m}\exp (-\mu _{1}\frac{m}{2})C_{0}(N,m)
\end{equation}
we get $F=-\frac{3}{2}\mu _{1}+c,c=3\sqrt{\frac{3}{2}}$, as for fixed $m$ we
have 
\begin{equation}
C_{0}(N,m)\sim \phi (m)N^{-\frac{5}{2}}c^{N}
\end{equation}
From (\ref{main}) the assertion of theorem 1 also follows in the
supercritical case, if we put $m=\beta N,0<\beta <1$ In the critical point
both expressions coincide, but the free energy is not differentiable at this
point.

To prove the second assertion of theorem 2 put $12\exp (-\mu _{1})=1+r$ and
estimate separately all 3 factors in (\ref{main}). We get, that there exists 
$0<\delta \ll \varepsilon \ll 1$ such that for any $m\leq \delta N$ 
\begin{equation}
\frac{P_{0,N}(m)}{P_{0,N}(\varepsilon N)}<(1+\frac{r}{2})^{-\frac{%
\varepsilon }{2}N}
\end{equation}
The result follows from this.

In the critical case the result follows similarly from from the estimates
(for even $m=\beta \sqrt{N}$) 
\begin{equation}
\frac{P_{0,N}(m+2)}{P_{0,N}(2)}\sim \prod_{k=1}^{\frac{m}{2}}(1+\frac{1}{2k}%
)(1-\frac{4}{3}\frac{k}{N})\sim C\sqrt{m}\exp (-\frac{1}{3}\frac{m^{2}}{N})
\end{equation}
Then for any $0<\alpha <\beta <\infty $ 
\begin{equation}
\lim (\frac{P_{0,N}(m(N)<\varepsilon \sqrt{N})}{P_{0,N}(\alpha \sqrt{N}%
<m(N)<\beta \sqrt{N})}+\frac{P_{0,N}(m(N)>\varepsilon ^{-1}\sqrt{N})}{%
P_{0,N}(\alpha \sqrt{N}<m(N)<\beta \sqrt{N})})=0
\end{equation}
as $\varepsilon \rightarrow 0$.

Let us remove now the coordinate system from the boundary. That is we do not
assume that homeomorphisms respect the origin (the specified edge) on the
boundary. The free energy remains the same. Only in the critical point the
distribution of the length changes - stronger fluctuations appear.

\begin{theorem}
In the critical point without coordinate system the boundary length is of
order $N^{\alpha }$ for any $0<\alpha <\frac{1}{2}$. The exact statement is
that the distribution of $\frac{\log m_{N}}{\log \sqrt{N}}$ converges to the
uniform distribution on the unit interval, that is $P_{0,N}(\frac{\alpha }{2}%
\leq \frac{\log m_{N}}{\log \sqrt{N}}\leq \frac{\beta }{2})\rightarrow \beta
-\alpha $ for all $0\leq \alpha <\beta \leq 1$.
\end{theorem}

From triviality of the automorphism group for almost all $T$, it follows
that $C(N,m)\sim \frac{1}{m}C_{0}(N,m)$. This gives in $f(N,m)$ the factor $%
1-\frac{1}{m}+O(\frac{1}{m^{2}})$ instead of $1+\frac{1}{m}+O(\frac{1}{m^{2}}%
)$ in the previous case. Similar calculations prove that asymptotically the
distribution coincides with the following family $\nu _{N}$ of probability
distributions on the set $\left\{ 1,...,\sqrt{N}\right\} $\ of $\sqrt{N}$\
elements 
\begin{equation}
\nu _{N}(i)=Z_{\sqrt{N}}i^{-1},Z_{\sqrt{N}}=\sum_{i=1}^{\sqrt{N}}i^{-1}
\end{equation}
It is easy to see that for $0\leq \alpha \leq \beta \leq 1$ 
\begin{equation}
\nu _{N}(\frac{\alpha }{2}\leq \frac{\log i}{\log \sqrt{N}}\leq \frac{\beta 
}{2})=\nu _{N}(N^{\frac{\alpha }{2}}\leq i\leq N^{\frac{\beta }{2}%
})\rightarrow \beta -\alpha
\end{equation}

\subparagraph{Conclusion}

We will make now some remarks concerning possible extension of these
results. In was known that in the pure gravity model without boundaries the
critical point itself is not universal: it strongly depends on the class of
triangulations. Similarly, the critical point $\mu _{1,cr}$ changes when we
consider similar models with $k$ holes. The corresponding critical points $%
\mu _{1,cr}(k)$ tend as $k\rightarrow \infty $ to the critical point for the
model without boundaries. Moreover, in the model without boundaries the
effect of fixing the origin changes the critical exponent from $-\frac{7}{2}$
to $-\frac{5}{2}$. A new phenomenon we proved in this paper is the
qualitative change of fluctuation scale.

The canonical ensemble we considered was inspired by topological field
theory, as a functor from $d$-dimensional to $(d-1)$-structures. Normally it
is introduced axiomatically \cite{ati}, algebraically (via Frobenius
algebras) or via path integrals. Discretization of path integrals is known
to be quite natural idea, see \cite{wata, ishka, baez2}. Here we considered
rigorously one of the possible scaling limits of it. We considered only one
hole case, some calculations for the case, when there are 2 or 3 holes
Anyway, this ensemble allows to give a rigorous existence proof of some
phases, observed earlier via other methods, see \cite{thorl}.

I would like to thank L. Pastur for useful discussions.

\end{document}